\documentclass[aps,twocolumn,superscriptaddress,showpacs,showkeys]{revtex4-1}
\usepackage[]{graphicx}
\usepackage{amsmath,amssymb,bm}

\begin{document}
\title{First-principles study on the stability 
of ($R$, Zr)(Fe, Co, Ti)$_{12}$ against 
2-17 and unary phases ($R$ = Y, Nd, Sm)}

\author{Taro \surname{Fukazawa}}
\email[E-mail: ]{taro.fukazawa@aist.go.jp}
\affiliation{CD-FMat, National Institute of Advanced Industrial Science
and Technology, Tsukuba, Ibaraki 305-8568, Japan}
\affiliation{ESICMM, National Institute for Materials Science,
Tsukuba, Ibaraki 305-0047, Japan}

\author{Yosuke \surname{Harashima}}
\affiliation{CD-FMat, National Institute of Advanced Industrial Science
and Technology, Tsukuba, Ibaraki 305-8568, Japan}
\affiliation{ESICMM, National Institute for Materials Science,
Tsukuba, Ibaraki 305-0047, Japan}
%\affiliation{Institute of Materials and Systems for Sustainability,
%Nagoya University, Nagoya, Aichi 464-8601, Japan}
\affiliation{Center for Computational Sciences, University of Tsukuba,
Tsukuba, Ibaraki 305-8577, Japan}
\affiliation{Division of Materials Science, Nara Institute of Science and Technology, Ikoma, Nara 630-0192, Japan}
\author{Takashi \surname{Miyake}}
\affiliation{CD-FMat, National Institute of Advanced Industrial Science
and Technology, Tsukuba, Ibaraki 305-8568, Japan}
\affiliation{ESICMM, National Institute for Materials Science,
Tsukuba, Ibaraki 305-0047, Japan}

\date{\today}
\begin{abstract}
The stability of ($R$, Zr)(Fe, Co, Ti)$_{12}$ with a ThMn$_{12}$ structure
is investigated using first-principles calculations.
We consider 
energetic competition with multiple phases that have the Th$_2$Zn$_{17}$ structure 
and the unary phases of $R$, Zr, Fe, Co, and Ti simultaneously by
constructing a quinary energy convex hull.
From the analysis, we list the stable phases at zero temperature,
and show possible stable and metastable ThMn$_{12}$ phases.
\end{abstract}
\pacs{TBD}
\preprint{Ver.0.1.1}
\maketitle

\section{Introduction}
%%%%%%%%%%%%%%%%%%%%%%%%%%%%%%%%%%%%%%%%
% Introduction
%%%%%%%%%%%%%%%%%%%%%%%%%%%%%%%%%%%%%%%%
Magnet compounds with the ThMn$_{12}$ structure 
have attracted attention as possible main phases
for high-performance magnets \cite{Ohashi88,Ohashi88b,Yang88,Verhoef88,
DeMooij88,Buschow88,Jaswal90,Coehoorn90,
Buschow91,
Sakurada92,Sakuma92,
Asano93,Akayama94,
Kuzmin99,
Gabay16,
Koener16,Ke16,Fukazawa17,Fukazawa19}.
Hirayama et al. recently reported that 
films of NdFe$_{12}$N, SmFe$_{12}$, and
Sm(Fe,Co)$_{12}$ have magnetic properties
superior to those of Nd$_2$Fe$_{14}$B,
which is the main phase of the current
best magnet.
%The rare-earths are 
%critical in obtaining high magnetic anisotropy,
%which is essential for magnets.
%
$R$Fe$_{12}$ ($R$ = rare earth) compounds with the ThMn$_{12}$ structure achieve high magnetization owing to the large Fe content.
However, the thermodynamic instability of these compounds is an obstacle for developing
practical magnets.

Introducing a third element can stabilize the 
structure of these compounds in the bulk material \cite{Ohashi87,DeMooij88,Coehoorn90}.
Ti is a typical stabilizing element, although it occupies the Fe sites in
$R$Fe$_{12}$, and thus
the magnetization is reduced due to the decrease in the Fe content.
Moreover, the magnetic moment of Ti couples antiferromagnetically to the host. 
In the search for an efficient stabilizer, 
several transition elements have been found to behave similarly to Ti.
$R$-site doping is another approach to stabilization, and
Zr is a potential stabilizer 
because it mainly dopes the $R$ site and simultaneous doping with Zr and Ti
may reduce the amount of Ti required for stabilization \cite{Suzuki14,Sakuma16,Kuno16,Suzuki16}. 
When the Ti content is small, the system tends to form structures related to 
the Th$_2$Zn$_{17}$- or Th$_2$Ni$_{17}$-structure (the 2-17 phase).
Y is another possible $R$-site doping stabilizer.
YFe$_{12}$ is also obtained by the rapid quenching method, and coexists with a
2-17 related phase that has the TbCu$_7$ structure \cite{Suzuki17}.
YFe$_{12}$ decomposes into the 2-17 phase and $\alpha$-Fe
at a higher temperatures.

Several theoretical studies have searched for efficient stabilizing elements.
We have investigated the stability and magnetism in NdFe$_{11}M$ 
for $M$ = Ti, V, Cr, Mn, Fe, Co, Ni, Cu, Zn by using first-principles calculations,
and we suggested that Co could function as a stabilizing third element \cite{Harashima16}.
%It should be noted that, however,
%we did not consider the stability against the 2-17 phase at this point.
The effect of the $R$-site substitution in $R$Fe$_{12}$ was also studied for 
$R$ = La, Pr, Sm, Gd, Dy, Ho, Er, Tm, Lu, Y, Sc, Zr, Hf \cite{Harashima18}.
In the paper, we considered the stability against the $R_2$Fe$_{17}$ phases and suggested
some rare-earth elements, including Y, as possible stabilizers.
However, how the proposed elements work as a dopant,
when the amount is fractional per unit cell, remains
theoretically unexplored.
The effects of co-doping have not been theoretically investigated for those systems either.
%One of the major difficulties was to keep energetic accuracy of calculation for non-stoichiometric systems
%with a feasible amount of resource consumption. 
%To overcome this point, we have developed a method for data integration between 
%approximated non-stoichiometric systems and precise stoichiometric systems \cite{Fukazawa19c}.
%We can take account of doping in target and competing phases using this method. 

In this paper, we consider the stability of ($R$,Zr)(Fe,Co,Ti)$_{12}$
to reveal how the dopant stabilizes the ThMn$_{12}$ structure.
We consider the unary phases ($R$, Zr, Fe, Co, Ti)
and the $R_2$(Fe,Co)$_{17}$ phases
with the Th$_2$Zn$_{17}$ structure as competing phases simultaneously.
In treating the randomness caused by the doping, we use 
coherent potential approximation (CPA).
We also use a previously proposed method
to improve the data from the CPA with a small number of accurate data from a different origin \cite{Fukazawa19c}.
We generate a quinary phase diagram
by constructing an energy convex hull,
and 
predict the stable phases at zero temperature within the scope of the calculation.
We also discuss the stability of the ($R$,Zr)(Fe,Co,Ti)$_{12}$ phases
based on their energy differences (hull distance) from the boundary of the convex hull
%he separate stable phase to decomposition,
and compare the values with previous experimental reports.

\section{Methodology}
%%%%%%%%%
% Methodology
%%%%%%%%%
We perform first-principles calculations based on density functional theory
within the local density approximation \cite{Hohenberg64,Kohn65}.
In solving the Kohn-Sham problem, we use AkaiKKR (MACHIKANEYAMA) \cite{AkaiKKR},
which is based on the Korringa-Kohn-Rostoker Green function method \cite{Korringa47,Kohn54}.
For the f-electrons in Nd and Sm, the open-core approximation \cite{Jensen91,Locht16,Richter98}  is used,
and the self-interaction correction is applied \cite{Perdew81}.
Calculations of
($R_{1-\alpha}$Zr$_\alpha$)(Fe$_{1-\beta}$Co$_\beta$)$_{12-\gamma}$Ti$_{\gamma}$ with
the ThMn$_{12}$ structure,
$R_2$(Fe$_{1-\beta}$Co$_\beta$)$_{17}$, and
Zr$_2$(Fe$_{1-\beta}$Co$_\beta$)$_{17}$
with the Th$_2$Zn$_{17}$ structure
are performed for
$\alpha$ from 0 to 1 with an interval of 0.1,
$\beta$ from 0 to 1 with an interval of 0.1, and
$\gamma$ from 0 to 1 with an interval of 0.5
($R$ = Y, Nd, Sm).
The lattice parameters are obtained by linearly interpolating
parameters from referential data.
These reference values are
obtained by first-principles calculations
for $R$Fe$_{12}$, ZrFe$_{12}$, $R$Fe$_{11}$Ti, ZrFe$_{11}$Ti, $R$Co$_{12}$,
$R_2$Fe$_{17}$, Zr$_2$Fe$_{17}$, $R_2$Co$_{17}$, and Zr$_2$Co$_{17}$
using QMAS, \cite{QMAS} which is based on the projector-augmented-wave method \cite{Bloechl94,Kresse99}.
%The energy values are improved by a previously proposed method \cite{Fukazawa19c}
%also using the referential data.

In treating the randomness caused by the doping, we use 
coherent potential approximation (CPA),
where electronic backscattering from the randomized environment is 
approximated by an effective medium which is self-consistently
determined in terms of the configuration average of 
the Green function.\cite{Soven67,Soven70,Shiba71}
However, calculating energy with the CPA tends to need much computational resources 
when high accuracy is needed owing to the use of localized bases in the CPA.
%In gathering data, therefore, we have to make a compromise in accuracy.
To overcome this problem, 
we use a recently proposed method for improving a data set with 
another data set from 
a different origin by constructing a stochastic model \cite{Fukazawa19c}.
In the current case, we apply this to the large data obtained with the CPA 
and a small amount of reference data, which are also
used in the interpolation for lattice parameters above.
%For the reference data, we obtained first-principles results
%for $R$Fe$_{12}$, ZrFe$_{12}$, $R$Fe$_{11}$Ti, ZrFe$_{11}$Ti, $R$Co$_{12}$,
%$R_2$Fe$_{17}$, Zr$_2$Fe$_{17}$, $R_2$Co$_{17}$, and Zr$_2$Co$_{17}$
%using QMAS, \cite{QMAS} which is code based on the projector-augmented-wave method \cite{Bloechl94,Kresse99}.

To investigate the stability, we construct an energy convex hull.
It is convenient to introduce
coordinates $\eta = (\zeta_1, \zeta_2, \zeta_3, \zeta_4, \varepsilon)$ for 
a quinary system
$R_{1-\zeta_1-\zeta_2-\zeta_3-\zeta_4}$Zr$_{\zeta_1}$Fe$_{\zeta_2}$Co$_{\zeta_3}$Ti$_{\zeta_4}$,
where $\varepsilon$ denotes the total energy per formula.
When there are two systems, which are $\eta^{(1)}$ and $\eta^{(2)}$ in this coordinate space, 
the composition and energy of
separate phases of $\eta^{(1)}$ and $\eta^{(2)}$ can be described as a linear interpolation
of these two points.
This concept can extend to phase separation into five different phases, which corresponds to
a five-dimensional facet that has five vertices, where a vertex represents a single phase.
In principle, the stable phases can be obtained by constructing facets from all the combinations 
of the vertices and
finding the energetically lowest facets.
The vertices of the lowest facets represent the stable phases.

We perform this part of the data analysis by using qhull, a package for efficiently constructing a convex hull
from given data points \cite{qhull,Barber96}.
This program also generates the upper part of a convex hull, which is unnecessary for our purpose,
although it is easy to remove by checking the direction of the normal vector for each facet.

\section{Results and discussion}
%%%%%%%%%
% Results and Discussion
%%%%%%%%%
\subsection{Stable phases}
% Stable phases
Tables \ref{tbl:Y--Zr}, \ref{tbl:Nd--Zr}, and \ref{tbl:Sm--Zr}
show the stable phases obtained by the hull construction
for $R$ = Y, Nd, Sm, respectively.
Ti-lean systems are of interest
because Ti-doping of ($R$,Zr)(Fe,Co)$_{12}$ reduces
the magnetization of a system by substituting Fe and Co.
The Zr-rich systems
without Ti are predicted to be stable.
There are differences in the range of Co content
for Zr(Fe,Co)$_{12}$ due to the differences in the lattice constants used in the calculations,
which are deduced by linear interpolation for each of the 
$R$ = Y, Nd, Sm cases.
However, the general trends are similar and 
the systems tend to be stable with a high Zr content 
and low Co content.
The estimated magnetizations of the stable Ti-less 
ThMn$_{12}$ phases
(shown in the captions of the tables)
are higher than the estimated value of 
1.49 T for Nd$_2$Fe$_{14}$B.\cite{Fukazawa19b}
\begin{table}
    \centering
    \caption{Stable phases in the Y--Zr--Fe--Co--Ti system.
    The calculated values of the magnetization are
    1.76 T for
    (Y$_{0.1}$Zr$_{0.9}$)Fe$_{12}$,
    and 
    1.74 T for Zr(Fe$_{0.9}$Co$_{0.1}$)$_{12}$.
    \label{tbl:Y--Zr}}
    \begin{tabular}{ll}
        \hline
        \hline
        Unary Y, Zr, Fe, Co, Ti \\ % #0--4
        \hline
        (Y$_{0.1}$Zr$_{0.9}$)Fe$_{12}$ \\ % #32
        % 1.753950        806.044000
        Zr(Fe$_{0.9}$Co$_{0.1}$)$_{12}$ \\ % #68
        % 1.735750        895.439000
        % Nd2Fe14B: 1.4915 1050 K
        \hline
        (Y$_{1-\alpha}$Zr$_\alpha$)Fe$_{11}$Ti \ ($\alpha = 0, 0.1, 0.2, 1$)\\ % #7,10,37
        (Y$_{1-\alpha}$Zr$_\alpha$)Co$_{11}$Ti \ ($\alpha = 0, 0.1, 0.2, 0.7, 0.8, 0.9, 1$)\\ % #337,340,343,358,361,364,367
        Y(Fe$_{1-\beta}$Co$_{\beta}$)$_{11}$Ti \ ($\beta = 0\operatorname{--}1$)\\
        %Y(Fe$_{1-\beta}$Co$_{\beta}$)$_{11}$Ti \ ($\beta = 0, 0.1, 0.2, 0.3, 0.4, 0.5, 0.6, 0.7, 0.8, 0.9, 1$)\\ % #7,40,73,106,139,172,205,238,271,304,337
        \hline
        Y$_2$(Fe$_{1-\beta}$Co$_{\beta}$)$_{17}$ \ ($\beta = 0,0.4,0.5,0.6,0.7,0.8,0.9,1$) \\ %#370,374,375,376,377,378,379,380        
        Zr$_2$(Fe$_{1-\beta}$Co$_{\beta}$)$_{17}$ \ ($\beta = 0,0.5,0.6,0.7,0.8,0.9,1$) \\ % #381,386,387,388,389,390,391
        \hline
        \hline
    \end{tabular}
\end{table}
%
% Table: Stable phases in Nd-Zr-Fe-Co-Ti
\begin{table}
    \centering
    \caption{Stable phases in the Nd--Zr--Fe--Co--Ti system.
    The calculated values of the magnetization are
    1.76 T for
    Nd$_{0.1}$Zr$_{0.9}$Fe$_{12}$
    and 
    1.74 T, 1.69 T, 1.64 T, 1.60 T, 1.54 T 
    for
    Zr(Fe$_{1-\beta}$Co$_\beta$)$_{12}$
    ($\beta = 0.1, 0.2, 0.3, 0.4, 0.5$),
    respectively.
    \label{tbl:Nd--Zr}}
    \begin{tabular}{l}
        \hline
        \hline
        Unary Nd, Zr, Fe, Co, Ti \\% #0--4
        \hline
        Nd$_{0.1}$Zr$_{0.9}$Fe$_{12}$ \\ % #32
        %  1.755060        847.915000
        Zr(Fe$_{1-\beta}$Co$_\beta$)$_{12}$ \ ($\beta = 0.1, 0.2, 0.3, 0.4, 0.5$)\\ % #68,101,134,167,200
        % 0.1: 1.736690        895.584000
        % 0.2: 1.689600        931.617000
        % 0.3: 1.643060        969.578000
        % 0.4: 1.596080        1016.172000
        % 0.5: 1.542280        1053.366000
        \hline
        Nd(Fe$_{1-\beta}$Co$_\beta$)$_{11}$Ti \ ($\beta = 0\operatorname{--}1$)\\
        %Nd(Fe$_{1-\beta}$Co$_\beta$)$_{11}$Ti \ ($\beta = 0, 0.1, 0.2, 0.3, 0.4, 0.5, 0.6, 0.7, 0.8, 0.9, 1$)\\ % #368,40,73,106,139,172,205,238,271,304,337
        (Nd$_{0.9}$Zr$_{0.1}$)Fe$_{11}$Ti \\ % #10
        ZrFe$_{11}$Ti \\ % #37
        (Nd$_{1-\alpha}$Zr$_{\alpha}$)Co$_{11}$Ti \ ($\alpha = 0.5,0.6,0.8,0.9,1$)\\ % #352,355,361,364,367
        \hline
        Nd$_2$(Fe$_{1-\alpha}$Co$_{\alpha}$)$_{17}$ \ ($\alpha = 0.5,0.6,0.8,0.9,1$) \\ % #375,376,378,379,380
        Zr$_2$(Fe$_{1-\alpha}$Co$_{\alpha}$)$_{17}$ \ ($\alpha = 0,0.9,1$) \\ % #381,390,391
        \hline
        \hline
    \end{tabular}
\end{table}
%
% Table: Stable phases in Sm-Zr-Fe-Co-Ti
\begin{table}
    \centering
    \caption{Stable phases in the Sm--Zr--Fe--Co--Ti system.
    The calculated values of the magnetization are 1.76 T for
    (Sm$_{0.1}$Zr$_{0.9}$)Fe$_{12}$
    and 
    1.74 T, 1.69 T, 1.64 T
    for
    Zr(Fe$_{1-\beta}$Co$_\beta$)$_{12}$ ($\beta = 0.1,0.2,0.3$)
    respectively.
    \label{tbl:Sm--Zr}}
    \begin{tabular}{l}
        \hline
        \hline
        Unary Sm, Zr, Fe, Co, Ti \\ % #0--4
        \hline
        (Sm$_{0.1}$Zr$_{0.9}$)Fe$_{12}$ \\ % #32
        % 1.755480        844.872000
        Zr(Fe$_{1-\beta}$Co$_\beta$)$_{12}$ ($\beta = 0.1,0.2,0.3$)\\ % #68,101,134
        % 0.1: 1.736440        895.412000
        % 0.2: 1.689390        931.758000
        % 0.3: 1.642820        969.709000
        \hline
        Sm(Fe$_{1-\beta}$Co$_\beta$)Ti ($\beta = 0\operatorname{--}1$)\\
        %Sm(Fe$_{1-\beta}$Co$_\beta$)Ti ($\beta = 0,0.1,0.2,0.3,0.4,0.5,0.6,0.7,0.8,0.9,1$)\\ % #368,40,73,106,139,172,205,238,271,304,337
        (Sm$_{0.9}$Zr$_{0.1}$)Fe$_{11}$Ti \\ % #10
        ZrFe$_{11}$Ti \\ % #37
        ZrCo$_{11}$Ti \\ % #367
        \hline
        Sm$_2$(Fe$_{1-\alpha}$Co$_{\alpha}$)$_{17}$ \ ($\alpha = 0.5,0.6,0.7,0.9,1$) \\ % #375,376,377,379,380
        Zr$_2$(Fe$_{1-\alpha}$Co$_{\alpha}$)$_{17}$ \ ($\alpha = 0,0.7,0.8,0.9,1$) \\ % #381,388,389,390,391
        \hline
        \hline
    \end{tabular}
\end{table}

\begin{figure}
    \centering
    \includegraphics[width=8cm]{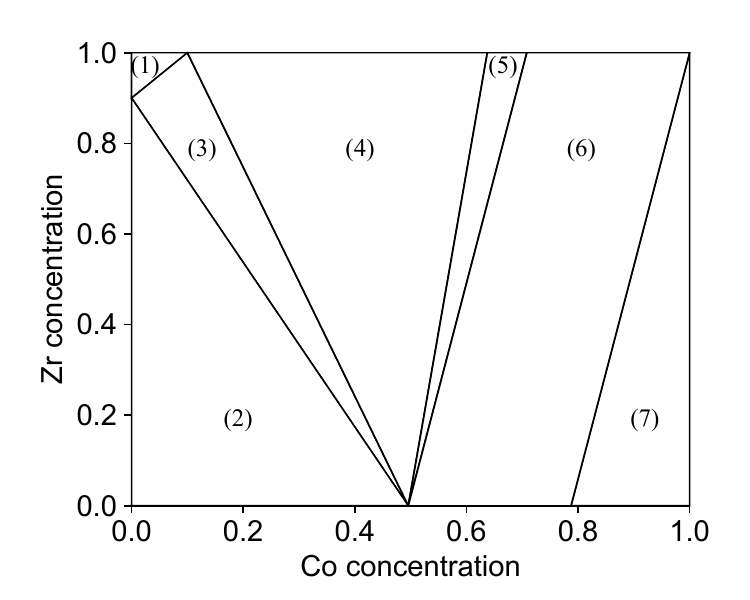}
    \caption{Phase diagram on the (Y,Zr)(Fe,Co)$_{12}$ plane.
    (1) (Y,Zr)Fe$_{12}$, Zr(Fe,Co)$_{12}$, Zr$_{2}$Fe$_{17}$, and unary Fe;
    (2) (Y,Zr)Fe$_{12}$, Y$_{2}$(Fe,Co)$_{17}$, and unary Fe;
    (3) (Y,Zr)Fe$_{12}$, Zr(Fe,Co)$_{12}$, Y$_{2}$(Fe,Co)$_{17}$, and unary Fe;
    (4) Zr(Fe,Co)$_{12}$, Y$_{2}$(Fe,Co)$_{17}$, Zr$_{2}$(Fe,Co)$_{17}$, and unary Fe;
    (5) Y$_{2}$(Fe,Co)$_{17}$, Zr$_{2}$(Fe,Co)$_{17}$, and unary Fe;
    (6) Y$_{2}$(Fe,Co)$_{17}$, Zr$_{2}$Co$_{17}$, and unary Fe and Co;
    (7) Y$_{2}$(Fe,Co)$_{17}$, Zr$_{2}$Co$_{17}$, and unary Co.
    \label{pd_YZr}}
\end{figure}

\begin{figure}
    \centering
    \includegraphics[width=8cm]{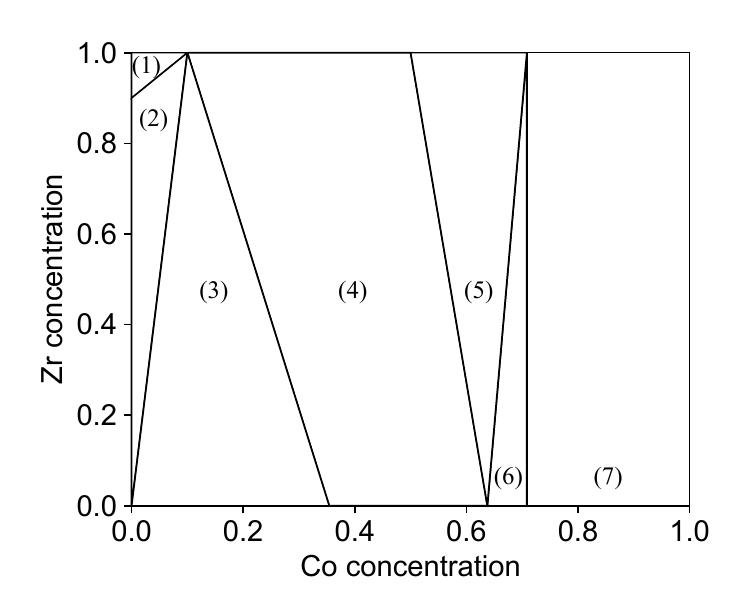}
    \caption{Phase diagram on the (Nd,Zr)(Fe,Co)$_{12}$ plane.
    (1) (Nd,Zr)Fe$_{12}$, Zr(Fe,Co)$_{12}$, Zr$_{2}$Fe$_{17}$, and unary Fe;
    (2) (Nd,Zr)Fe$_{12}$, Zr(Fe,Co)$_{12}$, and unary Nd and Fe;
    (3) Zr(Fe,Co)$_{12}$, Nd$_{2}$(Fe,Co)$_{17}$, and unary Nd and Fe;
    (4) Zr(Fe,Co)$_{12}$, Nd$_{2}$(Fe,Co)$_{17}$, and unary Fe;
    (5) Zr(Fe,Co)$_{12}$, Nd$_{2}$(Fe,Co)$_{17}$, Zr$_{2}$Co$_{17}$, and unary Fe;
    (6) Nd$_{2}$(Fe,Co)$_{17}$, Zr$_{2}$Co$_{17}$, and unary Fe;
    (7) Nd$_{2}$(Fe,Co)$_{17}$, Zr$_{2}$Co$_{17}$, and unary Fe and Co.
    \label{pd_NdZr}}
\end{figure}

\begin{figure}
    \centering
    \includegraphics[width=8cm]{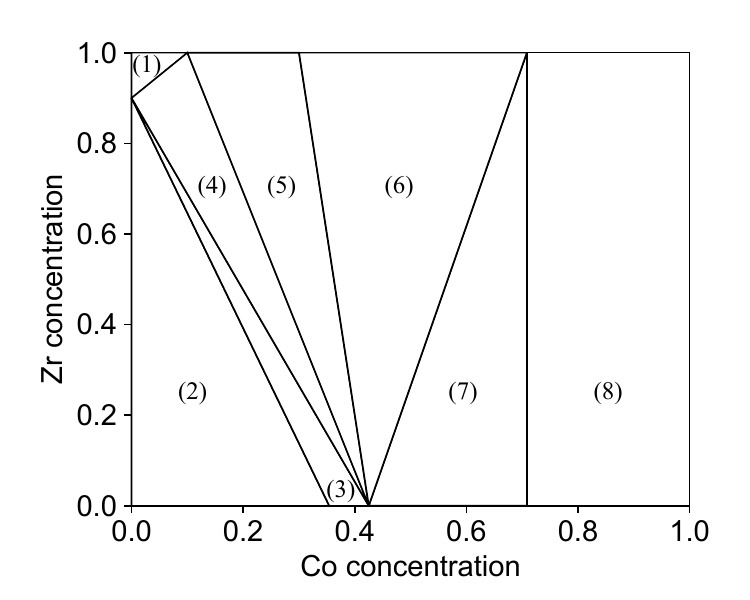}
    \caption{Phase diagram on the (Sm,Zr)(Fe,Co)$_{12}$ plane.
    (1) (Sm,Zr)Fe$_{12}$, Zr(Fe,Co)$_{12}$, Zr$_{2}$Fe$_{17}$, and unary Fe;
    (2) (Sm,Zr)Fe$_{12}$, Sm$_{2}$(Fe,Co)$_{17}$, and unary Sm and Fe;
    (3) (Sm,Zr)Fe$_{12}$, Sm$_{2}$(Fe,Co)$_{17}$, and unary Fe;
    (4) (Sm,Zr)Fe$_{12}$, Zr(Fe,Co)$_{12}$, Sm$_{2}$(Fe,Co)$_{17}$, and unary Fe;
    (5) Zr(Fe,Co)$_{12}$, Sm$_{2}$(Fe,Co)$_{17}$, and unary Fe;
    (6) Zr(Fe,Co)$_{12}$, Sm$_{2}$(Fe,Co)$_{17}$, Zr$_{2}$Co$_{17}$, and unary Fe;
    (7) Sm$_{2}$(Fe,Co)$_{17}$, Zr$_{2}$Co$_{17}$, and unary Fe;
    (8) Sm$_{2}$Co$_{17}$, Zr$_{2}$Co$_{17}$, and unary Fe and Co.
    \label{pd_SmZr}}
\end{figure}

%%%% Phase Diagrams
Figures \ref{pd_YZr}, \ref{pd_NdZr}, and \ref{pd_SmZr} show
the phase diagrams for the $R$ = Y, Nd, Sm cases, respectively.
These are obtained by projecting the lowest facets 
of the convex hull to the ($R$,Zr)(Fe,Co)$_{12}$ plane, that is, the plane of 
$\zeta_2+\zeta_3+\zeta_4 = 12/13$,
which constrains the ratio of Fe + Co + Ti to $R$ + Zr,
$\zeta_4 = 0$ (no Ti),
and $\varepsilon = \mathrm{const}$.
Therefore, these diagrams show how the 1-12 compositions
separate into stable phases at zero temperature.
%Figure: Phase Diagram Y--Zr
%
%Figure: Phase Diagram Nd--Zr
%
%Figure: Phase Diagram Sm--Zr

The ThMn$_{12}$ phases are found only in a limited region
of the diagrams, whereas the 2-17 phase appears
everywhere in the planes, except for the edges of regions, demonstrating how the 2-17 phase is important in considering
the stability of the ThMn$_{12}$ structure.

\begin{figure}
    \centering
    \includegraphics[width=8cm]{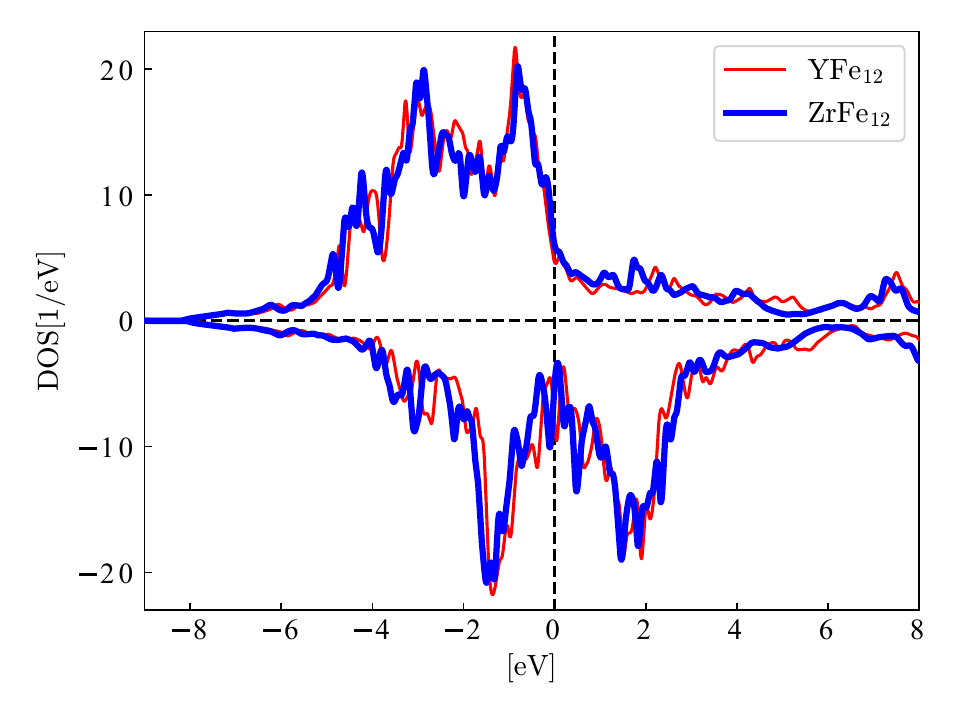}
    \includegraphics[width=8cm]{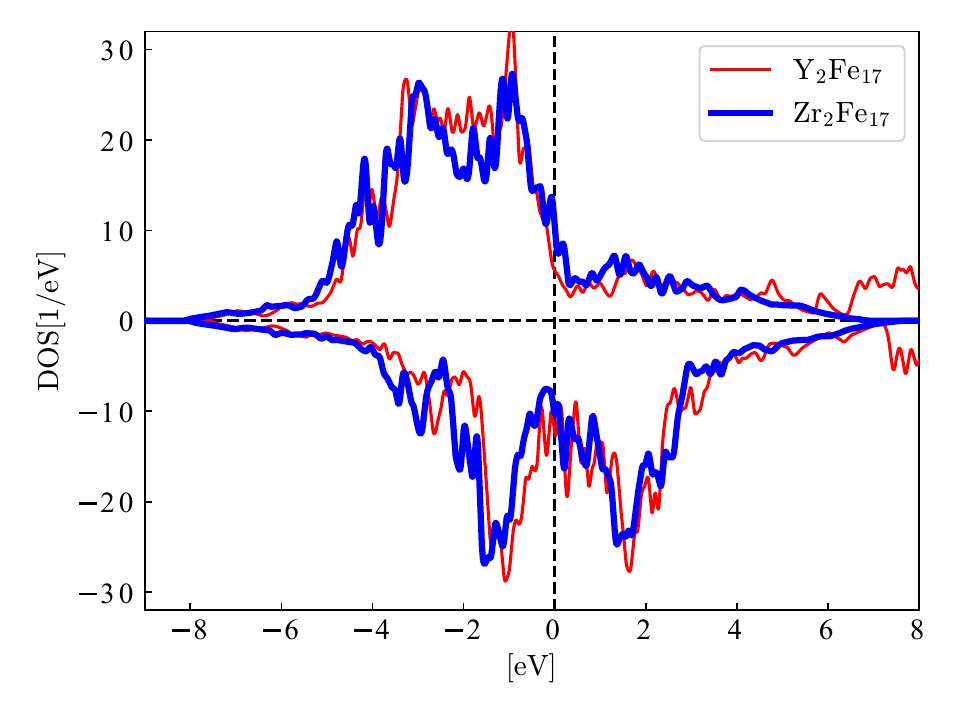}
    \caption{
        Total DOS of (top) YFe$_{12}$ and ZrFe$_{12}$
        and (bottom) Y$_2$Fe$_{17}$ and Zr$_2$Fe$_{17}$.
    \label{dos}}
\end{figure}

Related to the stabilization effect of Zr,
Friedel and Sayers discussed the role of 
d--d correlations in cohesion,
and discussed that
the cohesive energy tends to increase
when the number of electrons changes 
toward half-filling.\cite{Friedel77}
They also suggested that broadening of 
the band width is preferable for cohesion,
which is discernible in comparison between
the Y system and Zr system in Fig.~\ref{dos}.

The stabilization of the ThMn$_{12}$ phase
against the Th$_2$Zn$_{17}$ phase
by introducing Zr may be 
explained by the difference in 
DOS.
In ZrFe$_{12}$, 
the Fermi level is placed in a sharper dip in the minority band
than in Zr$_2$Fe$_{17}$,
and DOS at the Fermi level is close to 
that in Zr$_2$Fe$_{17}$. 
In terms of the band energy, the DOS of ZrFe$_{12}$
is more preferable under addtition and removal of 
a small amount of electrons, which corresponds to 
addition of Co and $R$). 

%No Ti0.5 systems
No system with $\gamma = 0.5$ is stable
in this search for 
($R_{1-\alpha}$Zr$_\alpha$)(Fe$_{1-\beta}$Co$_\beta$)$_{12-\gamma}$Ti$_{\gamma}$
stable phases.
Ti may be distributed inhomogeneously in separate phases of
the ThMn$_{12}$ phases with
$\gamma$ = 0 and 1, and the unary Ti phase.
However, this does not preclude the existence of a metastable phase with $\gamma = 0.5$.

\subsection{Hull distance}
% Hull distance
In this section, we examine the hull distance 
of %ThMn$_{12}$-type
($R_{1-\alpha}$Zr$_\alpha$)(Fe$_{1-\beta}$Co$_\beta$)$_{12-\gamma}$Ti$_{\gamma}$,
which is the energy difference from the stable separate phases.
We compare these results with experimental reports,
and roughly estimate the magnitude of deviation of the theory
in a manner similar to how Ishikawa et al.
estimated this for stoichiometric systems \cite{Ishikawa20,Ishikawa21}.

\begin{figure}
    \centering
    \includegraphics[width=8cm]{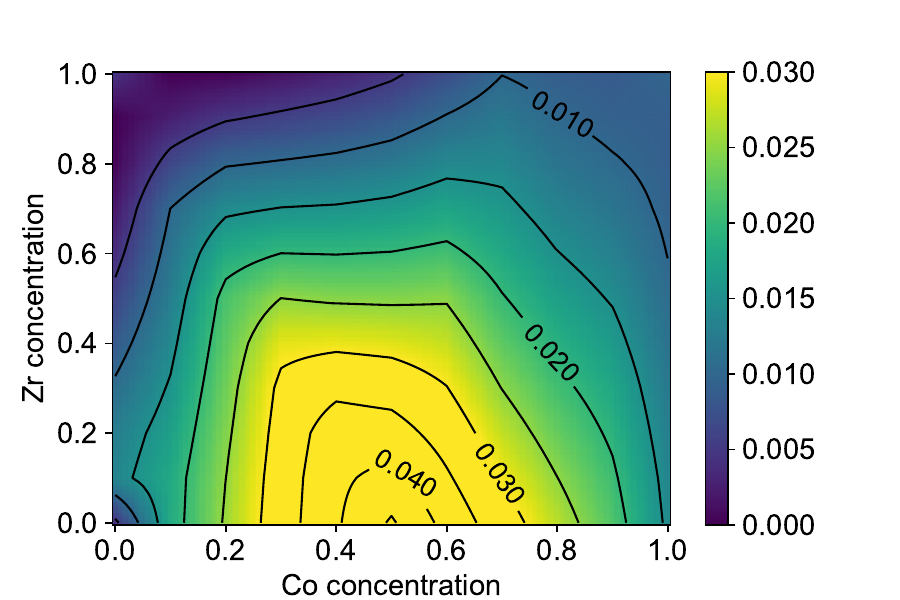} \\
    \includegraphics[width=8cm]{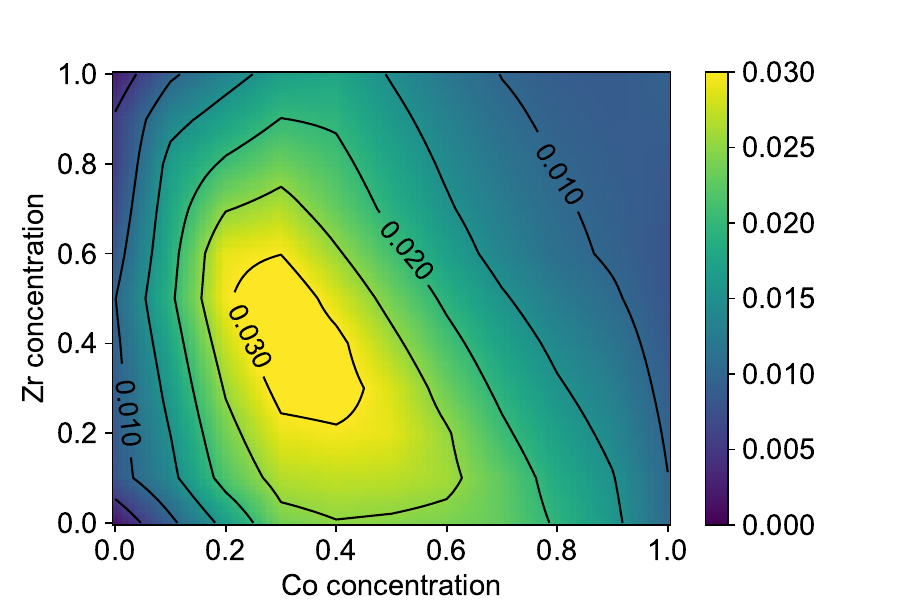} \\
    \includegraphics[width=8cm]{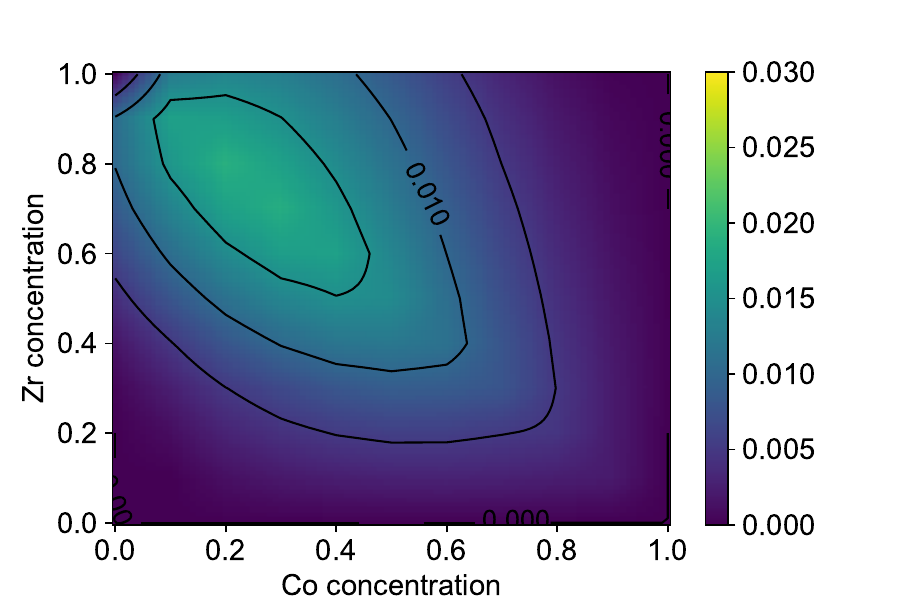}
    \caption{Hull distance in electronvolts for 
    (Y$_{1-\alpha}$Zr$_\alpha$)(Fe$_{1-\beta}$Co$_\beta$)$_{12-\gamma}$Ti$_{\gamma}$ with
    the ThMn$_{12}$ structure with 
    (top) $\gamma = 0$,
    (middle) $\gamma = 0.5$, and 
    (bottom) $\gamma = 1$.
    \label{hd_YZr}}
\end{figure}

\begin{figure}
    \centering
    \includegraphics[width=8cm]{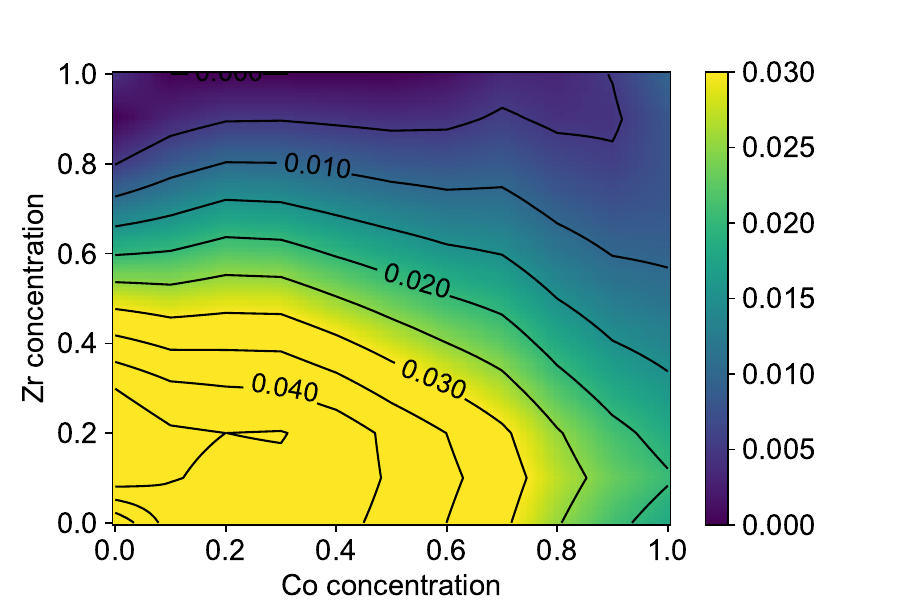} \\
    \includegraphics[width=8cm]{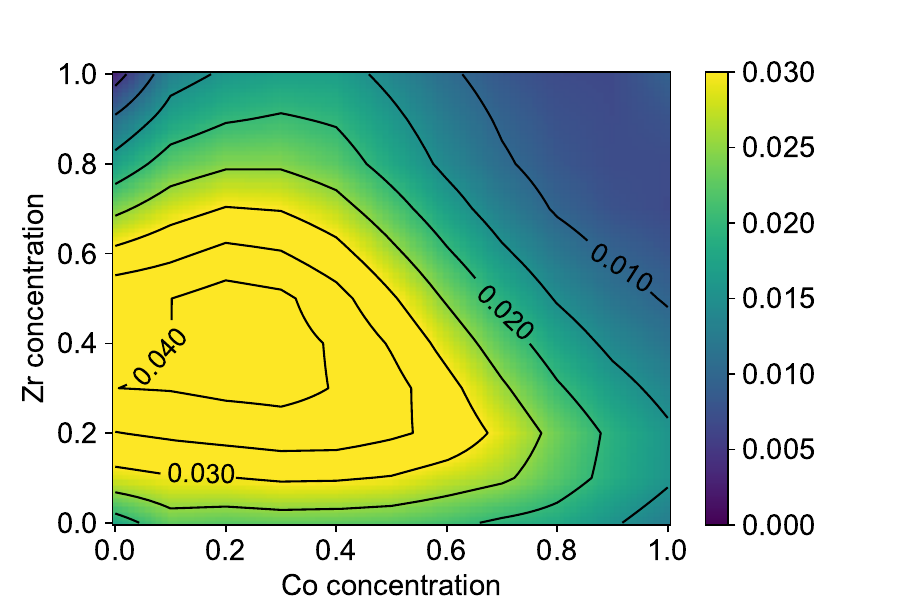} \\
    \includegraphics[width=8cm]{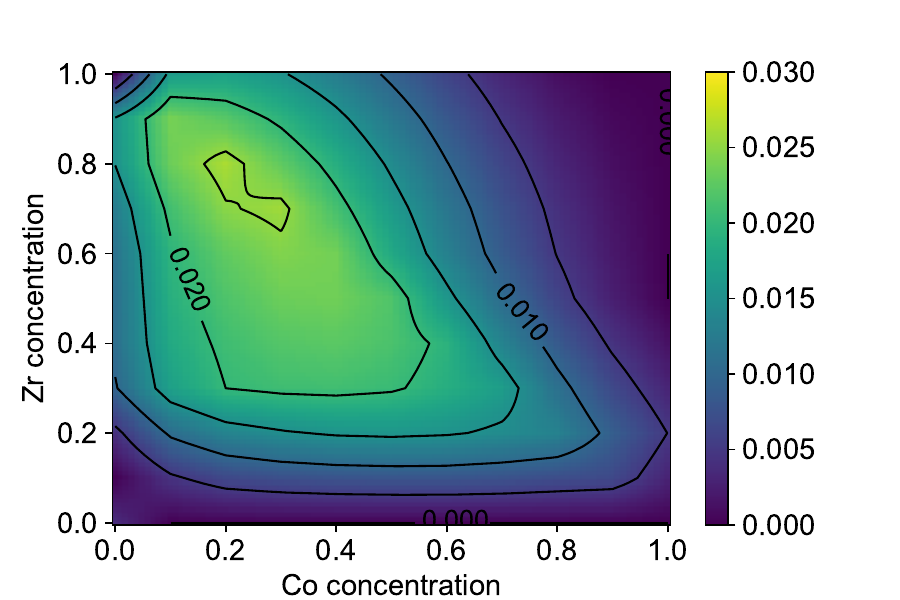}
    \caption{Hull distance in electronvolts for 
    (Nd$_{1-\alpha}$Zr$_\alpha$)(Fe$_{1-\beta}$Co$_\beta$)$_{12-\gamma}$Ti$_{\gamma}$ with
    the ThMn$_{12}$ structure with 
    (top) $\gamma = 0$,
    (middle) $\gamma = 0.5$, and 
    (bottom) $\gamma = 1$.
    \label{hd_NdZr}}
\end{figure}

\begin{figure}
    \centering
    \includegraphics[width=8cm]{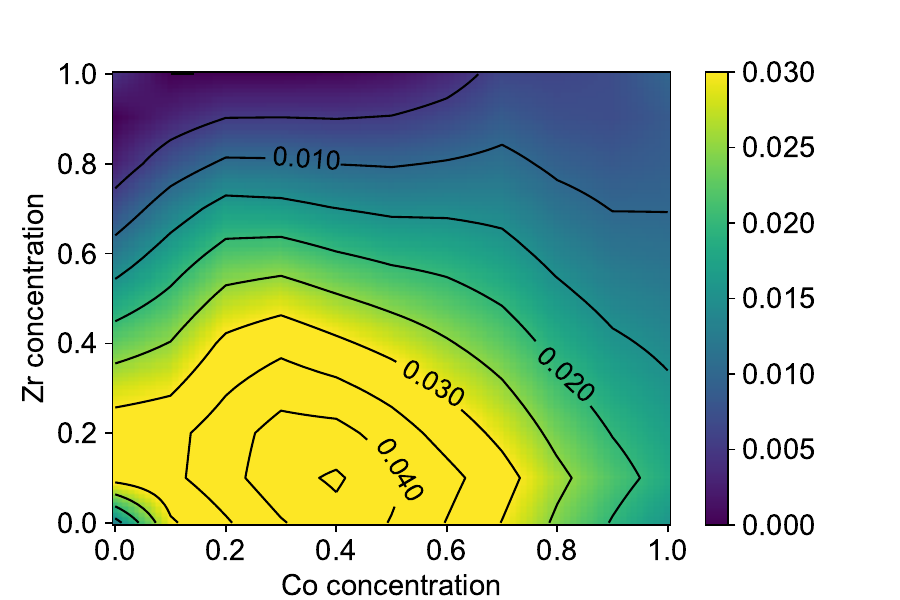} \\
    \includegraphics[width=8cm]{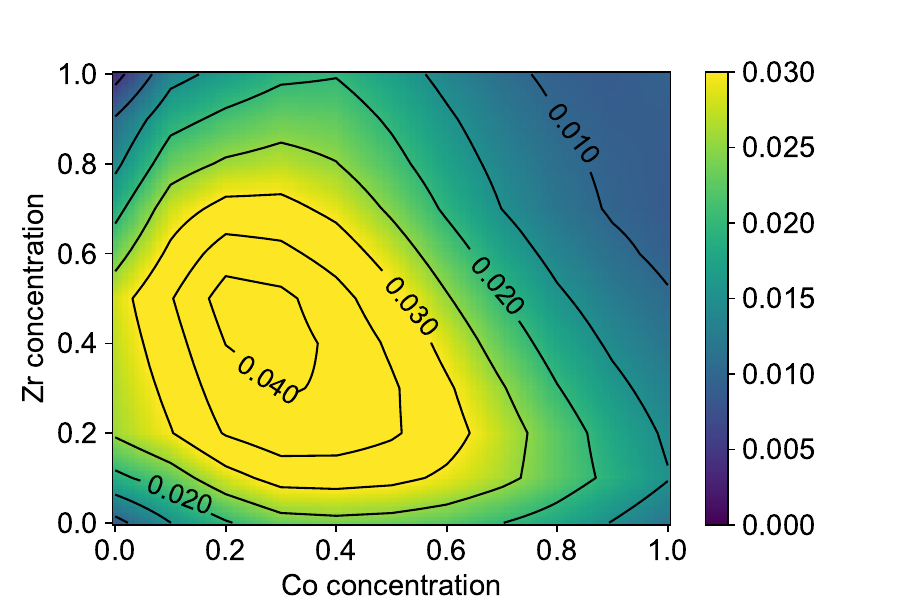} \\
    \includegraphics[width=8cm]{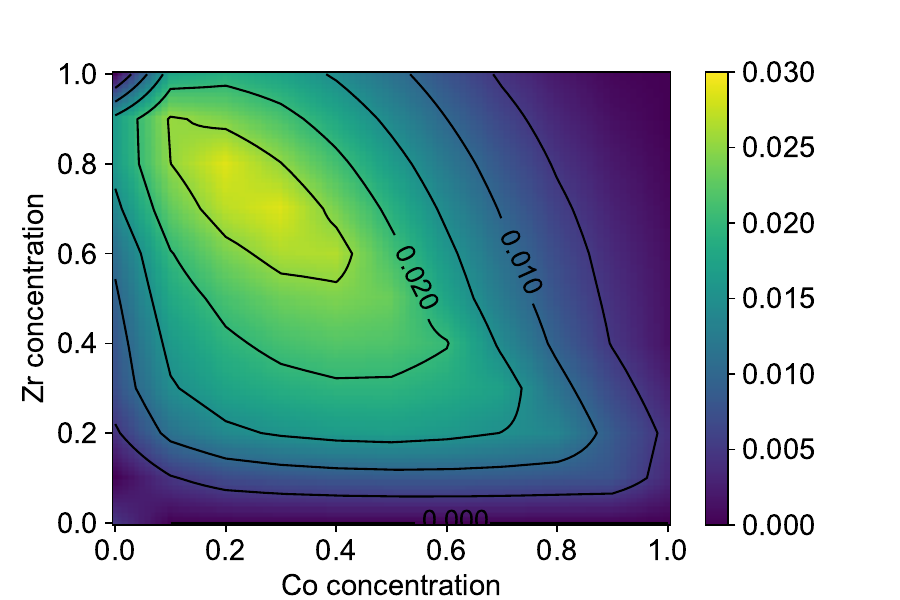}
    \caption{Hull distance in electronvolts for 
    (Sm$_{1-\alpha}$Zr$_\alpha$)(Fe$_{1-\beta}$Co$_\beta$)$_{12-\gamma}$Ti$_{\gamma}$ with
    the ThMn$_{12}$ structure with 
    (top) $\gamma = 0$,
    (middle) $\gamma = 0.5$, and 
    (bottom) $\gamma = 1$.
    \label{hd_SmZr}}
\end{figure}

Figures \ref{hd_YZr}, \ref{hd_NdZr}, and \ref{hd_SmZr} show the hull distance of
($R_{1-\alpha}$Zr$_\alpha$)(Fe$_{1-\beta}$Co$_\beta$)$_{12-\gamma}$Ti$_{\gamma}$
for the $R$ = Y, Nd, and Sm cases, respectively.
In each of the figures, the top panel shows results with $\gamma = 0$ (no Ti),
the middle with $\gamma = 0.5$, and the bottom with $\gamma = 1$.
Introducing Ti tends to decrease the hull distances of the ThMn$_{12}$ systems and the Co-rich and Zr-lean region.

Although the formation of ThMn$_{12}$ phases with compositions of
(Nd$_{0.7}$Zr$_{0.3}$)(Fe$_{0.75}$Co$_{0.25}$)$_{11.5}$Ti$_{0.5}$ and 
(Sm$_{0.8}$Zr$_{0.2}$)(Fe$_{0.75}$Co$_{0.25}$)$_{11.5}$Ti$_{0.5}$
by the arc-melt and strip-cast method have been reported \cite{Suzuki14,Suzuki16,Kuno16},
these compositions are in the peak of the calculated hull-distance function,
which suggests that those systems are energetically unfavorable.
Phase separation into stabler compositions may explain these results, and if this is the case, spatial inhomogeneity of elements is theoretically expected, which seems to 
be consistent with the microstructures observed experimentally by Suzuki et al. \cite{Suzuki14}
and Kuno et al. \cite{Kuno16}.

%Comparison with Hirayama films
Hirayama et al. have reported the epitaxial synthesis of films of
NdFe$_{12}$ \cite{Hirayama15,Hirayama15b} and
Sm(Fe$_{1-\beta}$Co$_{\beta}$)$_{12}$ ($\beta$ = 0, 0.1, 0.2) \cite{Hirayama17},
which had good homogeneity.
The calculated hull distances of the films were in the range of $\lesssim 30$ meV,
and this value could be used as a criterion for possible metastable phases.
Thus, we expect that it would be difficult to form a homogeneous system with 
a Co concentration within $0.3 \lesssim \beta \lesssim 0.7$, although
phases that are more Co-rich, such as $R$(Fe$_{1-\beta}$Co$_\beta$)$_{12}$ ($\beta \gtrsim 0.7$),
could be formed as a metastable phase.

\section{Conclusion}
%%%%%%%%%
% Conclusion
%%%%%%%%%
We examined the stability of 
($R$,Zr)(Fe,Co,Ti)$_{12}$ ($R$ = Y, Nd, Sm) with
the ThMn$_{12}$ structure based on first-principles calculations.
We considered energetic competition among the unary phases,
the (R,Zr)(Fe,Co,Ti)$_{12}$ phases, and
the $R_2$(Fe,Co)$_{17}$ and Zr$_2$(Fe,Co)$_{17}$ phases with
the Th$_2$Zn$_{17}$ structure.

The ThMn$_{12}$ structure was stabilized when the system
was more Zr-rich and Co-lean.
Zr stabilized the ThMn$_{12}$ structure in the framework
of the present calculation when the $R$ site was mostly substituted by Zr. 
Although it has been reported that 
Co can stabilize the ThMn$_{12}$ structure against the unary phases \cite{Harashima16},
we found that doping $R$Fe$_{12}$ alone with Co did not lead to a stable phase of 
the ThMn$_{12}$ structure because 
it stabilized the Th$_2$Zn$_{17}$ phase.

We calculated the hull distance of ($R$,Zr)(Fe,Co,Ti)$_{12}$ ($R$ = Y, Nd, Sm),
and discussed the stability of experimentally observed phases.
The uniform distribution of Zr appeared to be energetically unfavorable
from a theoretical perspective.
By comparing our results with an experimental report for
Sm(Fe$_{1-\beta}$Co$_\beta$)$_{12}$ ($\beta = 0, 0.1, 0.2$),
we identified a hull distance of 30 meV as a criterion for possible metastable states.
Provided this criterion is valid, 
formation of
Sm(Fe$_{1-\beta}$Co$_\beta$)$_{12}$ within the range of $\beta \gtrsim 0.7$
should be possible as a metastable state.

\section*{Acknowledgment}
This work was supported by 
a project (JPNP20019) commissioned
by the New Energy and Industrial Technology Development Organization (NEDO), 
% ESICMM
the Elements Strategy Initiative Center for Magnetic Materials (ESICMM, 
Grant Number JPMXP0112101004), 
% DPMSD (Fugaku)
and the ``Program for Promoting Researches on the Supercomputer Fugaku''
(DPMSD, Project ID: JPMXP1020200307) by MEXT.
% Supercomputers(ISSP, KUDPC)
The calculations were conducted in part using the facilities of the Supercomputer Center at
the Institute for Solid State Physics, University of Tokyo,
%and 
the supercomputer of the Academic Center for Computing and Media Studies (ACCMS), Kyoto University, 
% Fugaku
and the supercomputer Fugaku provided by the RIKEN Center for Computational Science
through the HPCI System Research Project 
(Project ID: hp200125, hp210179).

\appendix
%%%%%%%%%%%%%%%%%%%%%%%%%%%%%%%%%%%%%%%%
% Appendix A
%%%%%%%%%%%%%%%%%%%%%%%%%%%%%%%%%%%%%%%%
\section{Phonon dispersion in ZrFe$_{12}$}
\begin{figure}
    \centering
    \includegraphics[width=8cm]{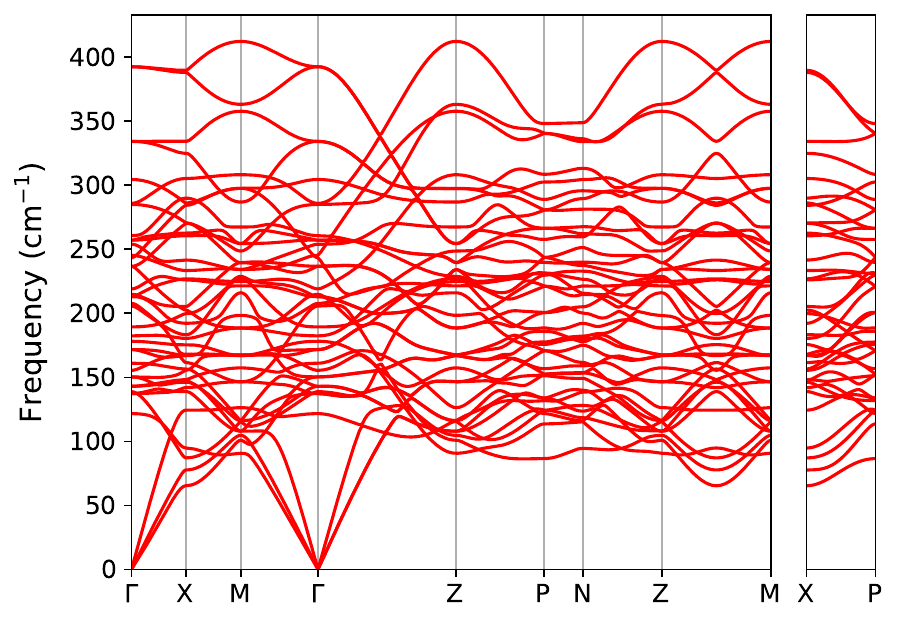}
    \caption{
        Phonon dispersion for 
        ZrFe$_{12}$.
    \label{phonon}}
\end{figure}

Some authors reported that they found 
no imaginary mode in $M$Fe$_{12}$ for 
$M$ = Dy, Y, Sm, Nd, Ce.\cite{Saengdeejing21,Xing21}
Although it is preferable to calculate such phonon
dispersions for all the systems predicted stable
in Tables~\ref{tbl:Y--Zr}--\ref{tbl:Sm--Zr}
to check the stability,
doped systems need much computational resource
in calculation.
We instead see ZrFe$_{12}$, which
has a close composistion to 
the stable Ti-less ThMn$_{12}$ phases,
and need only a small unit cell in calculation.
Figure \ref{phonon} shows calculated 
phonon dispersion for ZrFe$_{12}$ using 
the QMAS\cite{QMAS} and ALAMODE\cite{Tadano14}
package.
There found no imaginary mode in the dispersion.

\bibliography{seven}
\end{document}